# Poverty during Covid-19 in North Macedonia: Analysis of the distributional impact of the crisis and government response

Marjan Petreski

University American College Skopje

marjan.petreski@uacs.edu.mk

Abstract

In this paper we simulate the poverty effect of the Covid-19 pandemic in North Macedonia and we analyze the income-saving power of three key government measures: the employment-retention scheme, the relaxed Guaranteed Minimum Income support, and one-off cash allowances. In this attempt, the counterfactual scenarios are simulated by using MK-MOD, the Macedonian Tax and Benefit Microsimulation Model, incorporating actual data on the shock's magnitude from the second quarter of 2020. The results suggest that without the government interventions, of the country's two million citizens, an additional 120,000 people would have been pushed into poverty by COVID-19, where 340,000 were already poor before the pandemic. Of the 120,000 newly poor about 16,000 would have been pushed into destitute poverty. The government's automatic stabilizers worked to shield the poorest people, though these were clearly pro-feminine. In all, the analyzed government measures recovered more than half of the income loss, which curbed the poverty-increasing effect and pulled an additional 34,000 people out of extreme poverty. The employment-retention measure was regressive and pro-masculine; the Guaranteed Minimum Income relaxation (including automatic stabilizers) was progressive and pro-feminine; and the one-off support has been pro-youth.

**Keywords:** Covid-19, employment-retention scheme, relaxed Guaranteed Minimum Income support, one-off cash allowances, poverty, distributional effects

**JEL Classification:** I32, I38, D31

1. Introduction

As elsewhere worldwide, on March 11, 2020, the Macedonian government closed the school system in the country as a response to the outbreak of the novel coronavirus SARS-Cov-2, known as Covid-19. This was followed by the proclamation of the state of emergency on March 18th by the President of North Macedonia, which provided legal grounds[1] for the government to impose a curfew, a lockdown of high-contact sectors and reduced movement to the necessary minimum. A closure of the retailers selling non-essential goods soon followed, including bars, restaurants, cafes, and shopping malls. The strict phase of the lockdown continued until early-May 2020, when restrictions started to ease slowly and gradually. Although the government enacted stringent health protocols for the work of the high-contact sectors, consumers continued withholding their consumption, especially of non-essential goods. Travel remained considerably hard, particularly over the summer, when the Greek border – the dominant summer vacation outlet for Macedonians – remained closed. The middle of the summer saw increasing number of cases, which remained constant in the early autumn, and has only started accelerating recently (mid-October 2020). Hence, currently, the government is considering a new tightening of the restrictions of movement and behavior of the high-contact sectors, particularly after isolated cases of bars and clubs demonstrated negligence of health protocols.

The effects on the country's economy were massive, as even March saw significant declines in industrial production, trade volume and tourist inflow, let alone April and May, which exhibited dramatic deteriorations. It is projected that the economy will decline between 3.8% and 5.5% in 2020, while a drop as dramatic as the one in Q2-2020 of 1% has not been seen since the internal military conflict of 2001. Published data for Q2-2020 likewise suggest that at least a third of the sectors experienced a turnover decline exceeding 30% compared to the same period of the last year, while the decline in hours worked has been at about 25%. Employment declined more limitedly, due to the government job-retention measures, but perils persist as data from the agency of employment over the summer showed acceleration of unemployment filings. A rough estimation of the household income loss cannot be obtained from the official statistics (especially because the Survey on Income and Living Conditions covering the fiscal 2020 will be collected in the spring of 2021 and published only at the end of 2021). However, limited simulations exist: The World Bank (2020) estimates that the poverty in the country (calculated at the poverty threshold of 5.5 USD per day, PPP) will increase on average by 5 percentage points, while Finance Think (2020a) estimated a decline in disposable income in the magnitude of 7.2%. The World Bank (2020) relies on likely labor market impacts, by sector and employment type, before any government responses. Economic sectors of activity are grouped into (1) relatively unimpacted, (2) moderately impacted, and (3) highly impacted and arbitrary assumptions are used to simulate income losses in each of them.

The effect of the pandemic on poverty may be particularly aggravated in developing countries like North Macedonia. First, the lockdown affected tourism, trade and transport, which are usually low-pay sectors, i.e. including a large share of workers in or at the verge of working poverty. The same sectors have large shares of informal workers, as well seasonal ones. Atypical working contracts are likewise more frequent throughout the entire economy, particularly those with a definite duration, providing conditions where a job is easily lost, as well the likelihood of claiming unemployment insurance. Second, the country's system of social protection is well-functioning, but its sub-system of social assistance has been criticized for its poor targeting, resulting in an extreme poverty that still exceeds 4%. The comprehensive reform of June 2019 aimed to improve these deficiencies, but at the time the crisis hit, the reform had not yet reached its full potential. Third, according to official estimates, about 18% of the value added is created outside the formal economy, while recent IMF estimates suggest a share as high as 38%. A progressive income loss under Covid-19 will make the households in the informal economy poorer and more vulnerable. Fourth, the Macedonian economy heavily depends on remittances, whose amount is larger than that of FDI and official aid, and is in terms of living-standard effect larger than that of social assistance. A significant drop in remittances during Covid-19 will exert further deterioration of living conditions. An emerging literature (e.g. Patel et al. 2020) suggests that poverty is associated with higher risk of Covid-19 infection, as poorer households have bad housing conditions, and overcrowding. People in poverty are also more frequently employed in high-contact sectors, less often work in occupations that could switch to teleworking, and are more prone to health vulnerabilities, all providing lush grounds for contracting Covid-19, which may throw them into further destitute poverty.

---

[1] At the time Covid-19 hit, North Macedonia was preparing for the early elections then scheduled for April 12, 2020. Hence, the Parliament was deactivated, while the Government was only technical, i.e. with a mandate to conduct fair and democratic elections. The proclamation of the state of emergency was the only legal way to return the full law-making power to the Government, by enabling it to enact so-called Decrees with the power of law, hence bypassing the lack of parliamentary law-making power.

In order to cushion the effects of the lockdown, the government of North Macedonia introduced measures in four subsequent packages (of which the last one has been just announced and its deployment is expected early-November). As worldwide (see, e.g. OECD, 2020), measures have been roughly divided into income support, economic stimulus, and other measures. Economic stimulus measures include policies of interest rate cuts, deferred payments of tax advances by companies, several development loan packages with zero or subsidized interest, and the like. Other measures include halving of rents for tenants in state-owned apartments, halving of the penalty interest rates in paying public duties, as well a state-facilitated reprogramming of consumer loans, which secures that the principal is not repaid during the crisis months.

The income-support measures were aimed at preventing individuals from falling into unemployment and/or from facing severe income losses, and could be roughly divided into three groups. The first and the most massive measure has been the job-retention scheme, the disbursement of a direct financial support to companies who faced a drop in turnover over the critical months exceeding 30% annually, in the amount of the minimum wage per worker, with the requirement that the company would maintain the same number of workers two months after the subsidy finished. This measure included, firstly, a supplement in the form of subsidies of the company's social contributions to its employees in the amount of 50% in the hardest-hit sectors, namely: tourism, hospitality and transport. The second group encompassed social protection measures, whereby the primary intervention consisted of relaxation of the criteria for obtaining a guaranteed minimum income (GMI), primarily the proof of income based on the previous month rather than on the previous three, hence enabling a swift recognition of the income loss in order to enter the GMI system; and relaxation of the property-ownership criteria, whereby owing a house to live, a construction parcel of up to 500 sq.m. and a car older than five years was dropped from the exclusion list. The measures also included temporary relaxed criteria for obtaining an unemployment benefit, namely enabling it for persons who lost a job in March and April 2020, irrespective of the manner of contract termination, hence securing that those who resigned or who terminated the contract on a mutual agreement with the employer, which constituted ineligibility before, could temporarily enter the system. The third group of measures included one-off financial support to vulnerable citizens to aid consumption of domestic products and domestic tourism, in the range of 50 to 150 EUR for low-pay workers, the unemployed, and students.

The objective of the study is to nowcast the effects of the Covid-19 lockdown on poverty, i.e. the effects in terms of income loss, government budget, and distributional consequences. To judge these effects, the government income-support measures are crucial, and hence we also aim to assess their poverty and distributional effects. We go beyond aggregate estimates and quantitatively characterize the (income) groups most affected by the lockdown, identify who benefited from the government support measures and by how much, and the consequences in terms of poverty, inequality and the government budget. We relax and tighten the key assumption on the length of the effects of the pandemic onto the economy in order to judge the uncertainty surrounding our estimates. In this manner, we contribute to the emerging literature and ongoing debate on the yet uncertain effects of this unprecedented crisis on citizens and households, as well as if and how governments may be effective in rescuing jobs and incomes. Compared to limited existing estimations, this paper applies a more thorough and scientific approach, as well as considering an array of group-specific effects. What we refrain from in this paper is considering other and broader aspects of the crisis such as the reduced likelihood to get a job for those who were looking for it even before the pandemic, or the wider macroeconomic consequences, including the effect on public finance.

The results suggest that without the government intervention, Covid-19 would have resulted in about 11% of income lost in general, which soars to nearly 30% for the poorest households. The poverty-increasing effect of the pandemic is found to be sizeable, in the magnitude of a range of 6 p.p., to 25.8%, in terms of the upper absolute poverty rate (5.5 USD per day, PPP), resulting in an additional 120 thousand people of North Macedonia pushed into poverty (out of its population of about 2 million, of which 340,000 were already poor). The effect on extreme poverty (1.9 USD per day, PPP) is estimated at 0.8 p.p., to 4.5%, i.e. about 16 thousand additional people thrown into destitute poverty. Automatic stabilizers were found to have worked to shield the poorest (primarily located in the lowest two deciles), though they are clearly pro-feminine, as they reverse the extreme poverty-increasing effect in full.

All analyzed government-enacted measures to combat the economic and social effects of Covid-19 were important to reverse income declines. Slightly more than half of the lost income has been compensated through the measures, with the GMI relaxation and one-off support accruing more prevalently among the poorer households, while the employment-retention aid occurred along the entire income distribution, with the exception of the poorest households. With the measures, the decline of income in the poorest decile has been confined to about 8% the pre-pandemic level. This contributed to the curbing of the poverty-increasing effect, as measures shielded about 78 thousand people from falling into poverty. However, the ultimate effect on extreme poverty has been positive, as measures pulled an additional 34,000 out of it.

The study is organized as follows. Section 2 reviews the emerging literature relevant for the current pandemic of Covid-19. Section 3 presents the applied empirical methodology and the underlying data. Section 4 presents the results and offers a discussion. Section 5 concludes.

## 2. Emerging literature

There is abundant evidence in the literature of the impact of large macroeconomic shocks on societies. Most notably, with the case of the still recent Global Financial Crisis of 2007/8; Grusky et al. (2011) provide a comprehensive overview of the effects of this great recession on the economy, labor market, poverty and inequality. Still, the economic crisis of 2020 caused by the pandemic of the coronavirus Covid-19 is expected to dwarf the Global Financial Crisis of 2007/8 (Baldwin and di Mauro, 2020) in terms of the costs of health, human lives and widespread economic and social order. Pandemics, likewise representing large macroeconomic shocks, have been also examined with respect to their economic effects. Most notably, the current Covid-19 pandemic has been compared with the 1918 Influenza pandemic; for a comprehensive comparison, see Beach et al. (2020). Still, existing literature predominantly evaluates the 1918 Influenza effect on GDP growth, consumption and industrial production, corroborating its negative effect (Carillo and Jappelli, 2020; Dahl et al. 2020; Brainerd and Siegler, 2003). The evidence of the 1918 Influenza's effect on living standards has been scarce, mainly because of the lack of apposite data. However, limited information exists. For example, as mortality rates during the 1918 Influenza were high among the prime working-age group, the shrinking labor force increased wages (Garrett, 2009). Moreover, if poorer people die, the remaining earning cohort would be more equal, but this is hardly the case during Covid-19. Even during the 1918 Influenza, scarce evidence (e.g. Galletta and Giommoni, 2020, on Italian municipalities) suggests that the pandemic increased economic inequality through a reduction of the incomes at the bottom half of the income distribution.

The literature on Covid-19's effects on living standards, poverty and inequality is increasingly adding research. The International Labor Organization (ILO), the International Food Policy Research Institute (IFPRI) and the United Nations (UN) were the first to provide estimates on Covid-19 impact on poverty across the developing world. The ILO (2020) projected between 9 and 35 million new working poor (at the 3.2 USD line) in developing countries in 2020. The UN (Sumner et al. 2020) estimates more than 80 million new poor at the 1.9 USD line and 124 million at the 5.5 USD line, accentuating that most of them will be concentrated in Sub-Saharan Africa and South Asia. IFPRI (Vos et al. 2020) focus on Sub-Saharan Africa and South Asia and find 14-22 million people to have been pushed into extreme poverty (at the 1.9 USD line) due to the pandemic.

More thorough analyses focus on particular countries, although almost exclusively on developed ones. The key approach is microsimulation, as for the majority of the studies, actual data on the crisis's effects were not yet available in the months following its onset. Figari and Fiorio (2020) use a legislation-based approach to identify what occupations in Italy should be affected by the regulation. They rely on the EUROMOD – the European Tax and Benefit Microsimulation model-- and find that the poverty risk due to the Covid-19 lockdown is expected to increase over 8 percentage points. Beirne et al. (2020), likewise based on EUROMOD, consider arbitrary employment scenarios in Ireland and conclude that the disposable income of households is being reduced by the pandemic by as much as 20%, with the richest households suffering most, given their high pre-crisis employment probabilities. Along the same lines, O'Donoghue et al. (2020), studying Ireland, start from a scenario analysis for the sectoral shocks, and then distribute them based on an income-generation model. They corroborate Beirne et al.'s (2020) findings: income losses at the top of the income distribution were found to be eight times higher those of the bottom; moreover, the bottom 70% were found to have financially benefited from the government measures during the crisis, as opposed to the top 30%, who ended up worse off.

Bronka et al. (2020) rely on UKMOD – the British counterpart of EUROMOD-- whereby they impute the sectoral distribution of the Covid-19 shock they obtained from a dynamic IO-model parameterized with the results of a consensus analysis of 250 UK-based economists, and a probit model based on the Labor Force Survey to predict employment transitions within industry. Then, in UKMOD, they simulate the effect of the Covid-19 shock, the work of the automatic stabilizers, and the impact of the emergency measures put in place during the crisis. They find that the rescue package reversed the distributional impact of the pandemic, reducing poverty by more than a percentage point with respect to the pre-Covid-19 situation.

Han et al. (2020) construct new measures of the income distribution in poverty by relying on high-frequency data for a large, representative sample of American. families and individuals as an attempt to circumvent the problem with the data. They find that the government measures effectively countered the crisis effect on incomes, as poverty declined in the first months following the shock by 1.5 percentage points, despite employment dropping in April 2020 by 14%. They estimate that the Covid-19-related programs enacted by the U.S. government accounted for more than the entire decline in poverty, which otherwise would have risen by over 2.5 percentage points.

Lustig et al. (2020) is among the rare papers, to our knowledge, that performs a simulation exercise to estimate Covid-19's effects on poverty and inequality for developing countries in Latin America, including Argentina, Brazil, Colombia and Mexico. The study finds the worst effects of the crisis were overall concentrated in the middle class, while government responses largely offset the impact in Brazil and Argentina and less so in Colombia and Mexico. The analysis did not find a differentiated gender impact, despite the government measures being pro-women. However, Brazil showed that the incomes of the black population and indigenous people were impacted similarly to the incomes of the other segments of the population.

3. Methodology and data

3.1. MK-MOD Tax & Benefit Microsimulation model

Lack of longitudinal up-to-date information on household income and labor market circumstances, usually only available years after a shock occurred, imposes significant constraints for empirical analysis. Until recently in North Macedonia, economic data referent to the pandemic was almost inexistent. However, this has been slightly ameliorated with the publishing of a report on the value added, turnover and workers in the second quarter of 2020, when the impact of the crisis culminated. To address the limitations of this context, we assess the impact of the economic lockdown on household income by means of simulation of counterfactual scenarios by using a fiscal microsimulation approach (Figari et al. 2015).

We make use of MK-MOD, the EUROMOD-based tax-benefit microsimulation model for North Macedonia. Tax-benefit microsimulation models apply the fiscal and social-protection legislation to an observed input population, typically coming from survey data. This simulation utilized the Quality of Life Survey for MK-MOD. The most recent input data for MK-MOD is for 2017, however, we have inputted information from the national statistics to upgrade (i.e. reweight) the Survey to refer to the pre-Covid-19 situation. We use the tax-benefit microsimulation model to estimate how the welfare system protects people from an extreme shock, which became popular as a "stress test" of the tax-benefit system (Atkinson, 2009) and has been used to analyze the effects of the Great Recession (e.g. Jenkins et al. 2013).

MK-MOD is a static model where individual behavior (labor market activity, employment, childcare, saving, etc.) is assumed to be exogenous to the tax-benefit system. MK-MOD is a static microsimulation model which does not attempt to capture behavioral responses to policy changes. The model is based on theoretical considerations but is independent of any single theoretical perspective (Immervoll et al. 1999). In this setting, taxes and social transfers affect the labor-market behavior by changing the relative value of work vs. leisure, but MK-MOD is only a platform for adjacent analyses of behavioral change.[2] MK-MOD determines the different counterfactual scenarios in which we could identify individuals and households who are most likely to lose their income as a consequence of a shock. The simulated household disposable income depends on the cushioning effect of the automatic stabilizers (e.g. a declining market income automatically makes a household eligible for a guaranteed minimum income; or losing of a job makes an individual automatically eligible for an unemployment benefit). It addition, MK-MOD is set to be capable of capturing the effects of discretionary government policies implemented to shield sudden decline in income, as the ones implemented during Covid-19. MK-MOD has been validated by Petreski and Mojsoska-Blazevski (2017). Some further details of MK-MOD are provided in the Annex.

Given the objective of this study, in the empirical analysis we focus exclusively on income loss as one of the key channels through which Covid-19 pandemic affected individuals' well-being, and the subsequent cushioning of this effect by the government income-support measures. The overall effect of the pandemic, which is beyond the scope of this study, includes general equilibrium consequences and other behavioral responses (Figari and Fiorio, 2020). This closely relates to what MK-MOD can and cannot do.

Since MK-MOD is a static model, designed to calculate the first-round, immediate effects of a policy change, it neither incorporates the effects of behavioral changes nor the long-term effect of change. As such, it could be argued that MK-MOD is more suitable for analysis of local, relatively minor changes in taxes and transfers or in economic conditions – it is possible that its "out-of-sample" predictive power is limited. This may be particularly important for the objective of this paper, since the COVID-19 pandemic is certainly not a local or minor shock, and secondly it may imply sizeable behavioral effects and long-term scars. For example, based on the extent to which the pandemic affects individual and household incomes, individuals and couples may change their preference for work and leisure, with implications for the long-term household disposable income and work preferences. Another possibility would be that the design of the government measures to mitigate the socio-economic effects of Covid-19 may also exert long-term effects which MK-MOD presently cannot capture, a notable example being the exclusion of the high-paid jobs from the subsidy scheme, which may imply that the aggregate demand would suffer and, with it, the fate of low-income workers.

---

[2] MK-MOD has two adjacent models with behavioral component MK-Labor and MK-Pens, dynamic labor supply model and dynamic pension microsimulation model, respectively. See, e.g. Petreski and Petreski (forthcoming).

Certain caveats are related to the usage of the Quality of Life Survey (QLS). This is our own survey[3], collected in 2017 on a sample of 1,000 nationally representative households, comprising 4,071 individuals. The survey data was collected when the model was first built in 2017, but it has since been reweighted to reflect 2019 (the pre-pandemic situation). Namely, the existing weights were adjusted in a manner to reflect the trends in key indicators between 2017 and 2019; for example, reduction of unemployment between the two years implied that weights of unemployed were attenuated at the expense of the weights of employed. Weights for pensioners/elderly and children were likewise adjusted to reflect population dynamics. The usage of the Quality of Life Survey has some advantages and disadvantages compared to existing surveys of the national statistics (most notably, the Survey on Income and Living Conditions (SILC), on which such models are usually built). Namely, the QLS has been specifically designed to fit MK-MOD and hence asked questions which SILC does not ask, like the wage before retirement, length of tenure, household consumption (relevant for the simulation of indirect taxes), remittances, some indicators of wealth etc. Then again, QLS was based on only 1,000 households, as compared to the 5,000+ of SILC, hence facing larger margin of error, a byproduct of its lower cost. The cost has been the prime reason for the ad-hoc nature of the QLS as opposed to the SILC, which is annual, and hence the re-weighting we applied to reflect 2019. While re-weighting has not been dramatic, as changes in key indicators between 2017 and 2019 were far from large, re-weighting is still an "artificial" procedure which may create additional noise in the data. On the other hand, however, a transfer to SILC is almost impossible, as the survey is only available in the safe room of the national statistics office, on top of the fact that MK-MOD would require a serious recoding of variables (to fit a different survey structure), as well to deal with its segments which are possible through QLS but not through SILC.

A second data caveat relates to the simulation of the key government measure, which awards 14.500 MKD of subsidy to workers in companies who experienced a decline in revenues exceeding 30% in the most critical months of the pandemic compared to their average in 2019. Since the allocation of the measure is dependent on company-level information, we are unable to observe this in a household type of survey such as QLS. We approximate the allocation of funds from this measure by combining information from the national statistics on turnover in industry, trade and services in Q2-2020 with information on the sectors in which individuals work in our QLS. This certainly imposes noise in the calculation and should be borne in mind when interpreting the results.

Wits these caveats, we continue the modelling exercise in the manner in which MK-MOD has already been utilized.

---

[3] Collected for the purpose of designing MK-MOD back in 2017, by a professional private agency, yet by design and collection following all rules and procedures as done for the surveys of the national statistics.

### 3.2. Modelling the magnitude of the income losses

To model the effects of Covid-19, we apply a multilayer microsimulation approach. First, we define the counterfactual scenario based on the reweighted Quality of Life Survey 2017 to reflect the situation before the pandemic (2019). Second, we impute the income loss as observed in the second quarter of 2020. In the third step, we simulate the government income-support policies implemented in the period afterwards. Hence, we are simulating the chain of events between March and September 2020, which, given that the model is static, appear as if they occurred simultaneously (i.e. without a temporal component).[4] We explain the technicalities of the last two steps in detail.

To pursue the second stage of the simulation, we rely on the data made available by the State Statistical Office on the change in value added (or turnover), employment and hours worker per sector in the second quarter of 2020. Based on these, we classify the sectors as those highly affected (4) to those unaffected (1). As the government did not pursue any cuts in public wages, the public sector is by default classified as unaffected. Then, we make a couple of assumptions about how income behaved in the particular sectors, depending on how hard they were hit, according to the following scheme:

- For all sectors hit hardest (4), we assume that about 60 thousand jobs would have been lost (out of a total of nearly 600 thousand registered jobs) in absence of any job retention measure (a reflection of the two-month shutdown of the high-contact sectors). This number is derived from the evaluation of Finance Think (2020b) of the "14.500 MKD per worker" measure, reflecting a reduction of the wage mass of about 40%;

- For all sectors hit medium hard (3), we assume that all workers' salaries were reduced to minimum wage over a five-month period, and then their original wage was reduced by 30%;

- For all sectors hit medium (2), we assume that all workers reduced their wages by 30% over a five-month period, and then continued receiving their original wage;

- For all sectors not hit by the pandemic (1), we assume no changes in earnings.

For the self-employed, we assume the following scheme:

- For all sectors hit hardest (4), the enterprise was shut for five months, so that earnings ceased, while for the rest of the period income declined by 25%;

- For all sectors hit medium hard (3), we assume that income from self-employment declined by 50% over a five-month period, and then by 30%;

---

[4] Note that we assume that the tax evasion behavior did not change due to the pandemic, because this is something that is not explicitly modelled. The reality revealed contrasting evidence: on the one hand, there is limited anecdotal and largely unsupported evidence that retailers who remained open potentially increased the undeclared part of their turnover (to avoid VAT); while on the other, electronic payments increased and revenue inspections intensified, particularly after the "14.500 MKD per worker" measure was introduced. Hence, we do not have grounds to claim that tax evasion changed in a meaningful manner due to the pandemic.

- For all sectors hit medium (2), we assume that income from self-employment declined by 30% over a five-month period, and then reached its pre-pandemic level;
- For all sectors not hit by the pandemic (1), we assume no changes in earnings.

For those informally employed (i.e. those without formal contract), we assume income declined by 70%.

Although later we demonstrate a general resonance of these assumptions with the observed reality so far, they are still associated with significant uncertainty. Technically, this uncertainty is driven by the assumptions' construction (i.e. we bridge existing sectoral information from the national statistics from Q2-2020 with observable characteristics of employees in QLS). At the factual level, the assumptions are based on what we observed in the first three quarters of 2020, while the pandemic intensified in October 2020 (the time of writing of this paper), which implies that its effects may become worse than initially estimated. We therefore conduct a robustness check, whereby we relax the above assumptions to get a lower bound of the estimates, assuming that the worst effects of the crisis lasted for three instead of five months; and tighten the assumptions to get an upper bound of the estimates, whereby the crisis extends to nine instead of five months of 2020 (hence considering that the autumn wave of the pandemic hits hard). Such calculations would act as a confidence interval revealing the error pertinent to our estimates.

In this second step, we assume that the automatic stabilizer relevant for the GMI worked (despite that, in its original form, entry into the GMI program was allowed based on an assessment of a person's income over the preceding three months, but since we reduce a six-month period to a single static point, we assume that this automatic stabilizer produced immediate gains). This provides that whenever market income (here defined according to the wages from employment [incl. informal wages and self-employment) fell below the GMI threshold, the scheme saw new entrants without any rules having been changed. The same notion of thinking applies to all the other benefits and allowances which are means-tested, most notably child and educational allowances, parental allowances, and the like. Unfortunately, we cannot simulate the automatic stabilizer relevant for the unemployment benefit, for many reasons: i) it includes a strong temporal component, which we are lacking in the model; ii) eligibility is defined based on how the contract terminated (only those who were fired are eligible, but not those whose contract expired, who left of their own will or by a mutual agreement with the employer); iii) QLS did not ask about existing unemployment benefits. Regardless, the unemployment benefit – both the current scheme and the relaxation of the rules amid Covid-19 – is very low in terms of both spending and coverage and its inclusion would not affect the general conclusions of this study in a meaningful manner.

For taxes and social contributions, automatic stabilizers are allowed to work, i.e. the tax and social contributions revenues decline as jobs and incomes are lost. Pensions remain unaffected, reflecting the factual situation.

To complete the third part of the simulation, we assign a job-retention subsidy to the sectors to reflect the sectoral distribution of the scheme enacted by the Government of North Macedonia, as well the information derived from the sectoral turnover decline figures published in the state statistics. This was done to capture the condition of the measure that only companies who experienced a drop in revenues larger than 30% annually are eligible, according to the following scheme:

- All sectors hardest hit (4) closed during the critical months of the pandemic, but the job-retention government scheme secured that their workers would receive 50% of the minimum wage, provided legally with the 'force majeure' provision of the Labor Code, for three months. Furthermore, for the rest of the year these employees have continued to receive the minimum wage. This was because of the potential difficulties once the measure expired, where companies could not have introduced layoffs due to the subsidy's inherent design, stipulating that the number of workers must have been retained two months after the subsidy finished;

- In all sectors hit medium hard (3), wages were reduced by 25%, but not lower than the minimum wage over 3 months, and then were reduced by 20%;

- In all sectors hit medium (2), wages were reduced by 20% but not lower than the minimum wage over 3 months, and then returned to pre-pandemic levels;

For the self-employed, we assume the following scheme:

- For all sectors hit hardest (4), we assume that income from self-employment declined by 50% over a three-month period, and then by 20%;

- For all sectors hit medium hard (3), we assume that income from self-employment declined by 25% over a three-month period, and by 10%;

- For all sectors hit medium (2), we assume that income from self-employment declined by 10% over a three-month period, and then returned to its pre-pandemic level.

For the analysis of the social protection part of the government scheme, we rely on MK-MOD's properties to simulate entry into the GMI system. Namely, in addition to the work of the automatic stabilizer, which relates to the second stage of our simulation, we remove the property-ownership conditions for entry into the GMI system. Unfortunately, this is not very precise in MK-MOD and QLS, because relaxed conditions amid Covid-19 assume that the household is not in a possession of property unless there is a house in which its members live, a parcel with a building on it of up to 500 sq.m. and a car older than five years, which are rather tiny details that are not asked in standard surveys. Therefore, we approximate this by removing any property condition from the simulation of GMI.

The one-off support is modelled solely based on MK-MOD existing information, and particularly given that it was not conditioned on property ownership, but rather on labor market status and earnings.

What we do not model here is the behavior of any income from capital. On one side, we need to constrain our analysis here in a manageable framework. On the other, the simulation of income from capital (particularly of dividends) may be too speculative, as we are presently short of more succinct information of how it may have behaved under Covid-19. However, except for the richest segment of society, we do not expect that this constraint exerts any significant impact on our conclusions.

Likewise, we do not model any income stemming from remittances. The reason for not doing so is the weak coverage of remittances in our survey, as well the present lack of any guidance on the assumptions of how remittances behave at the individual/household level. The World Bank (2020) made an effort to incorporate the role of declining remittances in income loss during the pandemic, but the assumption of a proportionally equal decline of 40% across the income distribution seems unrealistic[5]. Nevertheless, the issue of remittances is very important for several reasons. First, Petreski and Jovanovic (2016) have shown that these are very important for the Macedonian economy, as their effect for the living standard is equal if not bigger than the effect of social assistance. Secondly, according to the World Bank[6], global remittances are expected to fall on average by 14% due to the pandemic. On the other hand, in North Macedonia they have already shrunk by 31% annually for the period March-June 2020, which is a remarkable decline. Given the role of informal channels for the transmission of remittances, we may reasonably doubt the number is exaggerated in times when such channels ceased to be used. Hence, it is not possible to draw a firm conclusion until we see how this unfolds by the end of 2020 at the earliest. However, it still points to the caution we need to impose in the usage of the simulations in this paper, as they do not consider the effect of changes in remittances, hence potentially underestimating the Covid-19 impact on income.

Based on the inputs defined in the second and the third step of the simulation, we arrive at the new sets of household income, based on which we calculate a set of poverty indicators: relative and absolute poverty rates (based on, respectively, the 1.9 USD and 5.5. USD per day poverty lines, PPP) and inequality indicators: the Gini coefficient and the decile distribution of incomes. We also derive the cost related to the government measures.

### 3.3. Income stabilization indicators

To measure the level of stabilization of incomes, we employ a set of measures also used in Figari and Fiorio (2020), as follows:

i) The net replacement rate (Immervoll and O'Donoghue, 2004), which measures the share of the post-disposable income of pre-disposable income:

$$NRR = \frac{Y_{post}^d}{Y_{pre}^d} = \frac{Y_{post}^o + B_{post} - T_{post}}{Y_{pre}^d}$$

Where: $Y_{post}^d$ is the household disposable income (original income ($Y_{post}^o$) plus benefits ($B_{post}$) minus taxes ($T_{post}$)) after the pandemic shock, and $Y_{pre}^d$, respectively, before the pandemic shock. Note that the original income is mainly composed of the market income from wages and self-employment, as well including some non-market components, like inter-household transfers and alimonies. Hence, under the pandemic shock, the original income may still be positive due to savings, inter-household transfers and the income from other household members. We do not simulate any changes in regard to the latter, since their coverage in QLS is quite feeble. I It would also be hard to assume their specific role and behavior during the crisis. Benefits are composed of the guaranteed minimum income, and child, educational, and parental allowances, while all the other allowances in the Macedonian social system, almost all of which are one-off, are taken as granted and are not simulated. The same applies for the unemployment benefit, for the reasons explained in Section 3.2.

---

[5] For example, Lustig et al. (2020) find (in the cases of Argentina, Brazil, Colombia and Mexico) that increases in poverty are worse than if one assumed proportionality in income declines.

[6] https://www.worldbank.org/en/news/press-release/2020/10/29/covid-19-remittance-flows-to-shrink-14-by-2021#:~:text=According%20to%20the%20World%20Bank's,of%203%20percent%20by%202030.

ii) The compensation rate (Salgado et al. 2004), which measures the extent of protection offered by government measures, i.e. the proportion of net earnings lost due to the economic lockdown compensated/rescued by the government measures net of taxes:

$$CR = \frac{(B_{post} - B_{pre}) - (T_{B_{post}} - T_{B_{pre}})}{Y^d_{pre} - Y^d_{post}}$$

Where notations are as before, and where the difference in disposable income before and after the shock represent income lost due to the lockdown, which is then compensated by government's net benefits. Taxes are allocated proportionally to each income source.

4. Results and discussion

4.1. The size of the income loss and budget effects: simulated versus available indicators

Before we embark on presenting and discussing the results, we need to verify income losses, to the extent possible, based on what we observed in reality. We first provide some insights into the potential income decline had there been no government intervention, according to sector at two digits of the NACE Rev.2. Namely, government support measures prevented declines in business income and/or hours worked from being reflected as declines in (workers') income. **Figure 1** presents the actual combination of drops in hours worked and turnover (an actual reflection of the crisis) and the simulated wage income declines (a simulated reflection of the crisis) for each of the two-digit sectors (we have data for 66 out of 89 sectors at this level). The regression line almost overlaps with the 45-degree line, suggesting that our simulation of wage income declines satisfactorily mimics the decline that would have happened had no government measures existed. About 45% of the variation in the simulated wage income decline is explained by the variation in the actual declines in hours lost and turnover declines, which reflects both a forecasting error and other factors that in reality affect wages and not hours/turnover, and vice versa (e.g. the share of the value added that goes to workers may differ in different industries). As we cannot split these two with the type and amount of information we have on disposal, we consider this validation as satisfactory.

Figure 1 – Comparison of actual and simulated losses in economic sectors

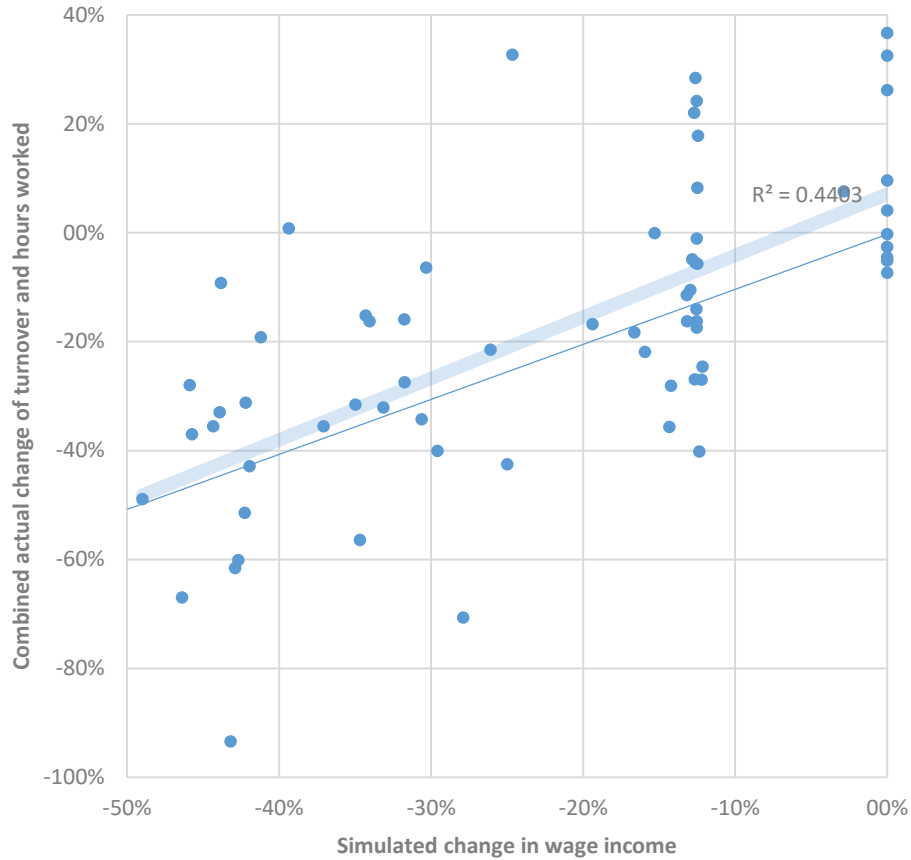

*Source: Author's calculations and State Statistical Office.*
*Note: The thin blue line is the 45-degree line.*

In the simulation, it is the hardest to capture the income rescue provided with the employment-retention measure "14.500 MKD per worker" for at least two key reasons. First, the measure was intended for workers in companies who experienced a decline in revenues exceeding 30%, in April and May 2020 compared to the average of 2019. However, we work with a survey of individuals, not companies, and hence this condition is not observed directly. Therefore, our assumptions set in section 3.2. should capture this condition only indirectly. Second, the measure provided income support for jobs over three months; however, its power also rested on the provision that the employer should maintain the same number of workers for at least two months after the support finished. Even after these two months elapsed, the incremental decline in the probability of laying off a worker may have been still higher than incremental increase in the probability of reemploying him had he been fired in a 'no-government support' scenario. This suggests that the 'income rescue' power of the measure extended beyond the actual government subsidy.

To get a sense of how our assumptions in section 3.2. pertinent to this measure reflect actual income loss, we conduct an exercise wherein we calculate the share of the income directly awarded to save jobs as three-fifths of the simulated income rescued (to reflect the fact that the income-saving power of the measure extended to at least five months, while the actual disbursement of the aid lasted for three months), factoring in the shares of formal workers per industry who were actually the only people entitled to the measure. Table 1 presents our simulations and contrasts them with the actual spending as per the announcement of the government institutions responsible for disbursement of the aid. The table shows that the direct income supported through this measure is fairly well-simulated, both in terms of level and in structure.

Table 1 – Budget effects of the key Covid-19 economic measures: simulated vs. actual

|  | | Simulated spending due to Covid-19 | Spending from administrative sources |
|---|---|---|---|
| | | *Million EUR* | |
| **Employment retention** | | | |
| | Manufacturing | 36.1 | 31.1 |
| | Trade | 13.2 | 14.6 |
| | Transport | 6.2 | 6.6 |
| | Hotels | 13.4 | 8.9 |
| | Recreation | 2.9 | 4.6 |
| | Rest | 18.0 | 12.7 |
| **GMI relaxation** | | 5.7 | 4.0 |
| **One-off support** | | 30.4 | 29.3 |
| | | | |
| Total spending | | 125.9 | 111.8 |

*Source: Author's calculations and various government institutions in charge for the deployment of measures.*

**Table 1** also presents the simulated budgetary effects of the other two measures we simulate here: the relaxed GMI conditions and the one-off support to consumption of the low-income individuals. Both simulated amounts fairly closely reflect the actual disbursements. The difference in the case of GMI relaxation may be due to the fact that the measure has still not reached its full power in reality, as it targets the poorest of the poor. These people may be facing information asymmetries and lack of documentation, which results in MK-MOD awarding them GMI status, while in reality they are actually not using this right. On the other hand, the one-off support has been automatic, resulting in smaller forecasting errors.

The total spending for the three key measures has been simulated at 126 million EUR, versus that of the actual one of 112 million EUR. We consider the difference to be a reflection of the above-discussed issues and, from that viewpoint, to be reasonable and acceptable.

### 4.2. Empirical evidence of replaced income and poverty

**Table 2** presents the poverty and inequality effects of the Covid-19 pandemic and the subsequent government measures. Column (1) presents the pre-Covid-19 situation, where relative poverty stands at 22.2% (2019). This rate closely mimics the 2018 rate published by the State Statistical Office of 21.9%, based on the Survey on Income and Living Conditions (SILC). The absolute poverty rates of 3.7% and 19.8% are only published by the World Bank and these for 2017 (the latest available year) are 4.6% and 19.5%. The difference appearing in the case of the extreme poverty may be a result of the fact that in our estimations for 2019 we already assumed the work of the social assistance system, reformed in 2019 to be taking effect. The last row presents the Gini coefficient, 31.8% in 2019, being almost the same (31.9%) as the one published by SSO for 2018 and based on SILC.

The effects of the crisis and the government measures are displayed one by one in columns (2) to (7), and we discuss them in detail below.

Table 2 – Poverty and inequality effects of Covid-19 and the government measures

|  | Pre-Covid-19 | Post-Covid-19 | | | | | |
|---|---|---|---|---|---|---|---|
|  |  | No intervention | | Measures | | | |
|  |  | Automatic stabilizers do not work | Automatic stabilizers work | Employment retention | Guaranteed minimum income relaxation | One-off support | TOTAL |
|  | (1) | (2) | (3) | (4) | (5) | (6) | (7) |
| Relative poverty (below 60% of the equiv. median income) | 22.2% | 24.5% | 24.3% | 23.7% | 24.4% | 24.1% | 23.4% |
| Absolute poverty, below upper middle income threshold (5.5 USD per day, PPP) | 19.8% | 25.8% | 25.8% | 22.9% | 25.8% | 25.3% | 21.9% |
| Absolute poverty, below extreme low income threshold (1.9 USD per day, PPP) | 3.7% | 4.5% | 3.8% | 4.0% | 4.1% | 3.1% | 2.0% |
| Income distribution (Gini) | 31.8% | 33.5% | 33.3% | 33.2% | 33.4% | 33.2% | 32.3% |

*Source: Authors' calculations based on QLS 2017 reweighted to reflect 2019 and MK-MOD.*

In a case of no government intervention and without the role of the automatic stabilizers (column 2), the pandemic would have resulted in severe income losses, further resulting in poverty hikes. The relative poverty would have increased by 2.3 percentage points (p.p.), to 24.5%, which is apparently not a large change, but is driven by the declining median income. That this is the case is corroborated by the absolute poverty rate (5.5 USD), which surges by a high 6 p.p. to 25.8%. This effect is aligned with the May 2020 calculations of the World Bank (2020). Likewise, extreme poverty increases by 0.8 p.p. to 4.5%. Hence, of the two million citizens, about 16,000 would have been thrown into extreme poverty, while about 120,000 would have fallen below the upper absolute poverty line, both representing a sizeable negative effect. This is a reversal of the gains in poverty alleviation attained since 2014.

Yet, the automatic stabilizers would have worked (column 3). We are constrained here to simulate only their effect on the recipients of social assistance, who are automatically entered into the system when their income decreases below a certain threshold, hence securing entry into the system. We are constrained to simulate the automatic stabilizer of the unemployment benefit, but this in North Macedonia is very stringent, as it captures only those who were fired by the employer (and leaves aside a very large contingent of those whose contract expired, who left on their own or where the contract's termination had been mutually agreed to). Hence, automatic stabilizers work only among the poorest segments, but their power is actually quite strong, as they compensate the effect the pandemic exerts on the poorest: extreme poverty rate reverts nearly to the pre-Covid-19 level. The 'no intervention' effect of Covid-19 results in an increase of income inequality by about 1.5 p.p., to 33.3%, which is a retraction of the gains attained in the last two years.

The next three columns present the incremental effect of the three key government measures. The employment-retention measure mainly works above the poorest segments (as the poorest are outside the labor market) and hence it results in improvements in relative poverty and upper absolute poverty. The positive effect it has on upper absolute poverty line is remarkable, as it removes almost half of the decline in poverty as would have been in a case where the measure had not existed. The effect on inequality is positive (in that it the measures decreased inequality) but is very small, suggesting that it acts along the income distribution. We should note that this effect may be underestimated, since we simulated through the sectoral impact, but were not able to exclude from the measure the workers whose wages exceeded 39.900 MKD net per month, as has been stipulated in the measure's design.

Column (5) presents the effect of the relaxed GMI criteria. It should be noted that the condition which stipulated that income is evaluated based on the previous month's income rather than on that of the previous three months is reflected in column (2), since our model is static and this is the only way to separate the two groups of relaxed criteria. The other criteria relaxation relates to the removal of most of the rules making property owners ineligible for GMI, and this is reflected in column (5). Its power is positive but weak, as the reduction of extreme poverty is only 0.4 p.p., to 4.1%. Yet, the overall effect of the relaxation of the GMI criteria could be loosely considered as the combination of columns (2) and (5), which for extreme poverty is in the magnitude of 1.1 p.p., suggesting that they are powerful to save about 22,000 from extreme poverty induced by Covid-19. We should accentuate that this is the theoretical maximum of the effect that can be observed, since we are not observing, nor could we capture, the behavioral responses of the new potential beneficiaries. This is particularly important as the measure targets the poorest of the poor, who usually face information asymmetries and are more prone to remain outside the social protection system despite being eligible for it.

Finally, the one-off support (column 6) primarily affected the poorest segment of society, although some positive effects are considered in other segments, as it targeted students in addition to low-pay workers and the unemployed.

The overall effect of the automatic stabilizers and government measures is provided in column (7). Compared to the pre-Covid-19 situation, the economy – with the amount of information we have at the time of writing of this paper – will end up with increased poverty. The relative poverty increase is estimated at about 24.000 people, while upper absolute poverty will reach about 42.000 people. These numbers are out of the total population of two million, of which 340,000 were already poor pre-pandemic. Extreme absolute poverty has declined, as about 34.000 people are dragged out of destitute poverty by a sizeable government intervention. In general, government measures and automatic stabilizers are powerful enough to save significant amounts of income at stake to be forgone permanently or for a protracted period of time. Another 24.000 people are prevented from falling into relative poverty, while 78.000 were prevented from falling into upper absolute poverty. The overall effect on the income distribution is slightly unfavorable (an increase of Gini by 0.5 p.p. to a level of 32.3%), yet a large part of Covid-19's devastating effects have been mitigated by the measures to this point.

**Figure 2** presents the uncertainties around our estimates presented in **Table 2**, mimicking confidence intervals. It should be recalled that such uncertainties are driven by two things: the yet unobserved full effects of the pandemic in 2020, and the manner in which we introduce the "14.500 MKD per worker" government measure in the model, since we do not observe companies but rather individuals. Two observations are relevant from **Figure 2**. First, the uncertainty is unevenly distributed around our point estimates, i.e. the risks for the living standard are not too detrimental, resulting in a higher upper bound for all three measures of poverty. Second, the poverty-reducing power of government measures is translated into reducing uncertainty around the estimates, which in a loose sense could be interpreted as the measures' power to somehow curb expectations among agents.

**Figure 2 – Uncertainties around the poverty estimates**

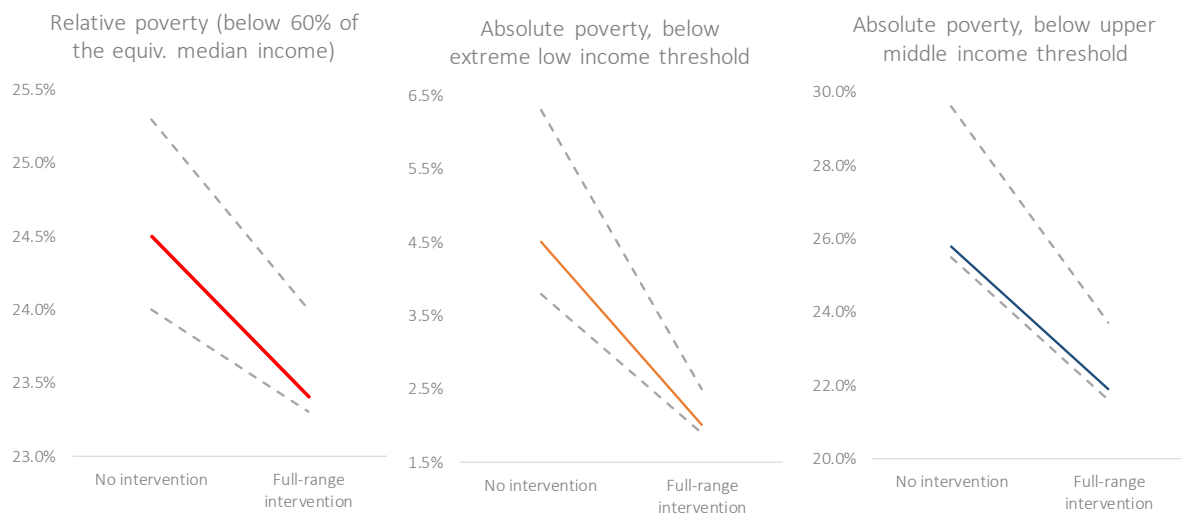

*Source: Authors' calculations based on QLS 2017 reweighted to reflect 2019 and MK-MOD.*

**Table 3** presents the poverty effects of Covid-19 for two critical groups: women and youth (ages 15-29). Women in North Macedonia represent a vulnerable group because they face lower employment and activation probabilities than men, and still earn less than men. Youth are considered vulnerable given their large unemployment rate (more than double the prime-age group) and resorting to atypical work contracts more frequently. Women faced a slightly higher poverty rate than men pre-pandemic (column 1), but seem to be slightly less hit by the pandemic (column 2). For them, the automatic stabilizers work more strongly, as extreme poverty is slightly lower than the pre-pandemic level (column 3). While this effect is more favorable than in the case of men, it is driven by the fact that women more frequently rely on social income than men. This is also reflected in the smaller rescuing power of the employment-retention measure (column 4), which works in the same fashion as for the entire population but the savings are of smaller magnitude for women. From that viewpoint, the employment retention measure could be considered pro-masculine. The GMI relaxation effect (column 5) is less also favorable for women. This is likely because men are more likely to own property and hence property-relaxation acted in a pro-masculine manner. However, when GMI relaxation is considered together with the work of automatic stabilizers, then the combined result is clearly pro-feminine. The one-off support, on the other hand, acts to reduce female poverty, but not as much as for the entire population, likely because of the prevalence of labor-market inactivity among women, where inactive labor force members are a category not captured by one-off support.

Table 3 – Poverty and inequality effects of Covid-19 and the government measures, for women and youth (15-29)

|  | Pre-Covid-19 | Post-Covid-19 | | | | | |
|---|---|---|---|---|---|---|---|
|  |  | No intervention | Measures | | | | |
|  |  | Automatic stabilizers do not work | Automatic stabilizers work | Employment retention | Guaranteed minimum income relaxation | One-off support | TOTAL |
|  | (1) | (2) | (3) | (4) | (5) | (6) | (7) |
| WOMEN | | | | | | | |
| Relative poverty (below 60% of the equiv. median income) | 23.1% | 24.2% | 24.2% | 23.6% | 24.2% | 23.8% | 23.3% |
| Absolute poverty, below upper middle income threshold (5.5 USD per day, PPP) | 19.9% | 25.7% | 25.6% | 22.9% | 25.7% | 25.2% | 21.8% |
| Absolute poverty, below extreme low income threshold (1.9 USD per day, PPP) | 3.4% | 4.4% | 3.3% | 4.3% | 4.2% | 3.1% | 2.1% |
| Income distribution (Gini) | 31.8% | 33.4% | 33.0% | 33.0% | 33.3% | 33.0% | 32.2% |
| YOUTH (15-29) | | | | | | | |
| Relative poverty (below 60% of the equiv. median income) | 28.0% | 30.3% | 30.1% | 28.9% | 30.3% | 29.8% | 28.6% |
| Absolute poverty, below upper middle income threshold (5.5 USD per day, PPP) | 23.5% | 30.7% | 30.7% | 27.5% | 30.7% | 30.0% | 26.5% |
| Absolute poverty, below extreme low income threshold (1.9 USD per day, PPP) | 3.6% | 5.2% | 3.9% | 4.6% | 4.5% | 2.7% | 1.9% |
| Income distribution (Gini) | 33.3% | 34.5% | 33.9% | 34.1% | 34.2% | 33.6% | 32.6% |

*Source: Authors' calculations based on QLS 2017 reweighted to reflect 2019 and MK-MOD.*

The results for youth are appealing. The observed patterns have the same general structure as before, though with distinct differential magnitudes. The Covid-19 crisis hit youth more than the overall population, as the increase in upper absolute poverty is by 7.2 p.p. (column 1). The automatic stabilizers reflect considerable shielding power, as extreme poverty declines by 1.3 p.p. to 3.9%, but the pre-pandemic level is not attained. Likewise, the employment-retention measure does not exert power as forceful as in the case of the overall population (where there is a poverty-reduction effect of only about 40%), and the GMI relaxation criteria also had a weak effect. It is likely that the one-off support provided the strongest poverty-reduction effect for youth, probably because it disbursed funds to the unemployed and students, a category in which youth are frequent participants Hence, the one-off support measure has been clearly pro-youth.

### 4.3. Distributional effects

The economic crisis caused by the coronavirus Covid-19 would have exerted an initial household income decline to about 89% of the pre-shock level, as revealed in **Figure 3**. The figure shows the net replacement rate by deciles along the income distribution. The original income after the shock declines to as much as 72% and 80% of the pre-Covid-19 level, in the two poorest deciles respectively, while in the richest decile it declines to only 98% of pre-COVID-19 levels (which could be underestimated as we do not model income from capital; see section 3.2).

Figure 3 – Net replacement rates along the income distribution

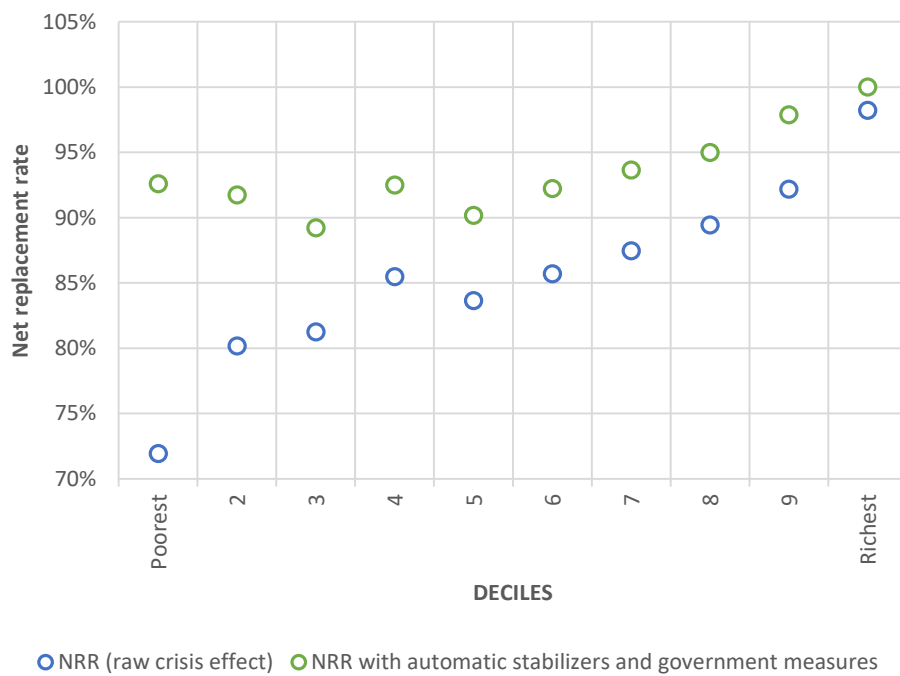

*Source: Authors' calculations based on QLS 2017 reweighted to reflect 2019 and MK-MOD.*

To focus on the income protection offered by the government measures, we adopt the compensation rate, presented in **Figure 4**. It shows that the average net public contribution to disposable household income as a proportion of net earnings lost due to the pandemic is 53.8%, with a pattern along the income distribution revealing a decline in on the lower end of the distribution and an increase on the higher end. Expectedly, the bulk of the government support is channeled through the employment-retention measure, which is spread along the income distribution, although with clear regressivity. Again, it must be noted that such a distribution may be slightly overestimated (where the regressivity pattern may be similarly overestimated), as we were not able to exclude the jobs with salaries exceeding 39.900 MKD per month (about 4% of all jobs). In any case, we argue that regressivity should not be of a concern in such unusual times, because jobs on the higher end of the income distribution are important for the aggregate demand and hence for the medium-term impact of the crisis on economic growth. Households in the poorest decile, expectedly, benefited the least from the employment-retention measure, since they are predominantly outside the labor market and instead rely on social income. For them, as well for the second-lowest decile, automatic stabilizers (in the case of GMI) play a strong role in the prevention of poverty, as well the one-off support, which is yet also distributed throughout the other deciles, mainly due to its component to financially support students in addition to low-paid and unemployed individuals.

Figure 4 – Compensation rates along the income distribution

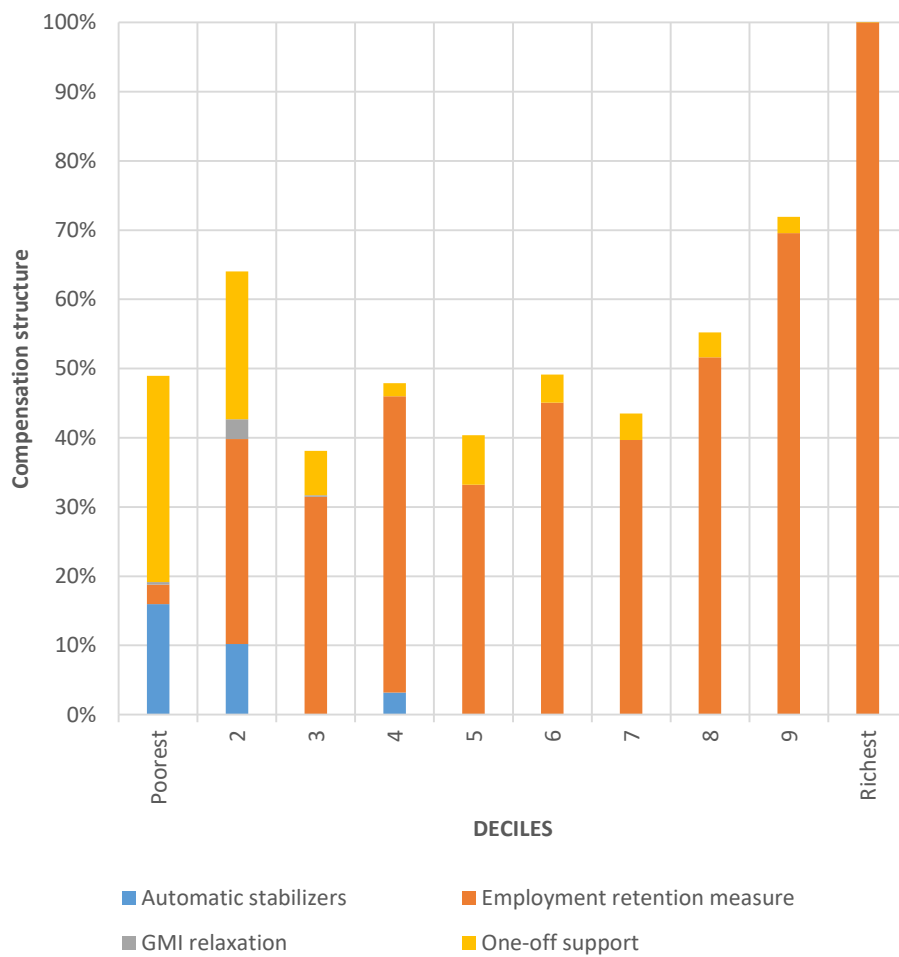

*Source: Authors' calculations based on QLS 2017 reweighted to reflect 2019 and MK-MOD.*

With this compensation, the average post-Covid-19 household income climbed to as high as 95% of its pre-crisis level, with the largest gain accumulated in the poorest decile, and with the size of the compensation then declining along the income ladder.

## 5. Conclusions and policy implications

In this paper we simulated the effect that the economic and social crisis caused by the outbreak and spread of the coronavirus has had on poverty in North Macedonia. We also analyzed the extent to which the Macedonian social protection system provides income support for those affected by the lockdown at the onset of the pandemic. Finally, we provided estimates of the income-saving power of three key government measures adopted after the Covid-19 outbreak. The paper offers a scenario rather than a forecast and it provides a reference point against which one could evaluate new policies. In this attempt, counterfactual scenarios are simulated by using MK-MOD, the Macedonian Tax and Benefit Microsimulation Model – the Macedonian counterpart of EUROMOD, which is the EU-wide microsimulation model, integrated with information from sectoral developments in value added, turnover, employment and hours worked for the second quarter of 2020, recently published by the national statistical office of North Macedonia.

The key results of the counterfactual analysis are as follows. Without government intervention, Covid-19 would have resulted in about 11% of income lost in general, which soars to nearly 30% for the poorest households. The poverty-increasing effect of the pandemic is found to be sizeable, in the magnitude of 6 p.p. (poverty increased to 25.8%) in terms of the upper absolute poverty rate (5.5 USD per day, PPP), resulting in additional 120,000 North Macedonians becoming poor (of its two-million population, of which 340,000 were already poor pre-pandemic). The effect on extreme poverty (1.9 USD per day, PPP) is estimated at 0.8 p.p., being increased to 4.5%, or about 16,000 additional people pushed into destitute poverty. Automatic stabilizers were found to have worked to shield the poorest (primarily located in the lowest two deciles) from these effects, though are clearly pro-feminine, as they reverse the extreme poverty-increasing effect in full.

All analyzed government-enacted measures to combat the economic and social effects of Covid-19 were significant when it came to rescuing losses in income. Slightly more than half of the lost income has been recovered through the measures, with the GMI relaxation and one-off support occurring more prevalently among the poorer households, while the employment-retention aid occurred along the entire income distribution, with the slight exception when it came to the poorest households. With the measures, the decline of income in the poorest decile has been confined to about 8% the pre-pandemic level. This helped to curb the poverty-increasing effect, as measures saved about two thirds of it (counted in terms of upper absolute poverty), i.e. they prevented about 78,000 people from falling into upper absolute poverty. While the ultimate effect on extreme poverty has been positive, as it declined to only 2%, i.e. not only prevented 16,000 individuals from being pushed into extreme poverty, it also dragged additional 34,000 out of it. The employment-retention measure has been pro-masculine, as men face higher employment rates than women; GMI relaxation (including automatic stabilizers) has been pro-feminine, as women are less active on the labor market and more often rely on social income; the one-off support has been pro-youth, as the unemployed and students constituted two crucial parts of the measure's target group.

The analysis and its conclusions have several policy inferences. First, the design of the key government measures seems to have been appropriate from the perspective of saving income affected by the pandemic. The distributional effects are likewise favorable, in the sense that measures shielded income in a progressive manner, i.e. they protected the income of mainly the poorest and less so of the richest. However, this may have been driven by the pattern of the crisis itself: it hit poorer segments significantly more than the richer ones. However, second, there might have been sub-groups which were hit hard but were not helped through government relief. One notable example is those who, despite the employment-retention measures, lost their jobs. The plight of these individuals was addressed through a one-time temporary unemployment insurance expansion for the period March-April 2020, but job loss intensified thereafter. These people may also not fulfill the GMI criteria since a majority of them are not the poorest of the poor, nor do they qualify for the stringent condition of the unemployment benefit (with its pre-pandemic design). One way to shield these vulnerable-to-poor individuals is to relax the criteria for the unemployment benefit as a pre-step for a longer-term reform of the unemployment insurance system. The other way to shield them would be to launch some type of temporary social aid, which would not be conditional on anything except income, hence securing fast entry into the system with a well-communicated feature of temporality to help these individuals bridge tough times, while also securing that they return to the labor market once the pandemic subsides. Temporary aid may be also important for children living in poor households, because they represent the other segment where no direct government measure has been deployed. Petreski et al. (2020) find that government measures were in general pro-children, but only indirectly, through recovering income losses among poorer segments of society, who usually have more children on average. However, along Petreski et al.'s (2020) suggestion, relaxed GMI criteria may be replicated in the case of child allowance, to provide easier access to the benefit for children in families who lost their income during the crisis.

In terms of the employment-retention measure, we documented its strong ability to reduce absolute poverty, but we also documented its regressive structure, i.e. awarding in absolute terms more to the richer segments (which, on the other hand, prevents relative poverty from increasing significantly, through affecting the median income). This is not a surprise in the sense that employed individuals are not concentrated among the poor segments (being one apparent reason why they are poor). Despite the fact that we did not precisely capture the exclusion criterion for this measure by leaving aside all high-paid jobs (about 4% of all employed), the regressivity of the measure is still apparent. On the one hand, awarding a subsidy to jobs which faced difficulties but which were unlikely to have been lost is a clear deadweight loss and should be certainly avoided. On the other hand, it is questionable whether tight and exact targeting was possible at the onset of the crisis when the world faced many unknowns. Secondly, the support of the better-paid jobs in difficult times may be much more important than in normal times, as good job matches imply value to the economy (beyond the employer's and employee's self-interest), and avoiding the destruction of these matches probably has positive effects in the long run that could not be captured by the model. Therefore, the future redesign of this measure should be thought of in terms of who has been hardest hit by the crisis and not in terms of whether the job was high-paid or low-paid. The October 2020 announcement of the redesign of the measure may have been pointed in that direction, but only its delayed deployment will reveal the true result.

Finally, one-off support may be worthwhile to be considered as an ad-hoc instrument in the future again, especially if the effects of the crisis linger, but caution is needed in two regards. First, the measure must have a clear income test, even if targeting sub-populations generally known as poor and vulnerable (e.g. single parents). Second, if repeated, the measure must clearly acknowledge that it is really one-off, because repeated one-offs would create expectations that the aid would be provided again. From that viewpoint, a temporary aid scheme may be considered for a certain period of time (e.g. 6 or 12 months), covering the period of the second wave of the pandemic (autumn 2020) and its potential aftershocks in the spring of 2021. Even in this case, the income test and the temporality of the measures must be clear at the outset, particularly in order not to generate a political-economy cost when the measure is withdrawn.

The reader should note that this analysis abstracts from any possibility of income and consumption smoothing that households may exploit over longer periods of time. As Figari and Fiorio (2020) argue, individual preferences for consumption smoothing lead to a decrease in current consumption in the presence of economic insecurity, so that in a case of no government intervention, the overall effects of the crisis may be significantly exacerbated. Finally, some additional caveats emerging in the literature should be borne in mind: the analysis does not say anything about health inequalities inflicted by the Covid-19 crisis, nor on how these may interfere with income inequality (see, e.g. Baker, 2019). Inequality during health emergencies in Covid-19 should not be overlooked in general, especially given that those who are more exposed to the health risk are also those belonging to disadvantaged socioeconomic groups. From that viewpoint, the fact that the employment-retention scheme is regressive in nature (in relative terms, it awards more to the higher-income segments, despite that the absolute amounts may be proportionally smaller) should be considered in any attempts to redesign the measure. On the other hand, this is only limitedly addressed by the progressivity of the GMI relaxation measure, inter-alia because the latter is more than ten times cheaper than the employment-retention scheme. As funding of the government measures is related to heavy borrowing by the government during Covid-19 – accumulating to nearly 10% of GDP – such considerations will certainly come to the fore when a discussion to introduce progressivities in the taxation system reopens post-Covid-19.

## References

Atkinson A. B. (2009) Stress-Testing the Welfare State. In: B. Ofstad, O. Bjerkholt, K. Skrede and A. Hylland (eds), Rettferd og Politik Festskrift til Hilde Bojer, Emiliar Forlag, Oslo, 31-39, 2009.

Baker, C. (2019) Health inequalities: Income deprivation and north/south divides. House of Commons Library, 22 January 2019 (available at https://commonslibrary.parliament.uk/insights/health-inequalities-income-deprivation-and-north-south-divides/). [25 October 2020]

Baldwin, R. and W. di Mauro, B. (Eds.) (2020) Mitigating the COVID economic crisis: Act fast and do whatever it takes. London: Centre for Economic Policy Research.

Beach, B., Clay, K. and Saavedra, M.H. (2020) The 1918 Influenza Pandemic and its Lessons for Covid-19. NBER Working Paper No. 27673.

Beirne, K., Doorley, K., Regan, M., Roantree, B., and Tuda, D. (2020) The Potential Costs and Distributional Effect of Covid-19 Related Unemployment in Ireland. ESRI Series, Budget Perspectives 202101.

Brainerd, E. and Siegler, M.V (2003), The Economic Effects of the 1918 Influenza Epidemic. Available at SSRN: https://ssrn.com/abstract=394606.

Bronka, P., Collado, D. and Richiardi, M. (2020) The Covid-19 Crisis Response Helps the Poor: the Distributional and Budgetary Consequences of the UK lock-down. EUROMOD Working Papers Series, EM11/20.

Carillo, M.F. and T. Jappelli (2020) Pandemics and Local Economic Growth: Evidence from the Great Influenza in Italy. Centre for Studies in Economics and Finance (CSEF), University of Naples, Italy.

Dahl, C.M., C.W. Hansen and P.S. Jense (2020) The 1918 epidemic and a V-shaped recession: Evidence from municipal income data. *Covid Economics*, 6.

Figari, F., A. Paulus and H. Sutherland (2015) Microsimulation and Policy Analysis. In: Handbook of Income Distribution Volume 2B, edited by A. B. Atkinson and F. Bourguignon, Elsevier.

Figari, F., and Fiorio, C. (2020) Welfare resilience in the first month of Covid-19 pandemic in Italy. EUROMOD Working Papers Series, EM6/20.

Finance Think (2020a) To what extent will Covid-19 increase poverty in North Macedonia? Policy Brief 43.

Finance Think (2020b) Was it necessary to provide financial support to companies to retain jobs during the Covid-19 crisis? Policy Brief 41.

Galletta, S. and T. Giommoni (2020) The Effect of the 1918 Influenza Pandemic on Income Inequality: Evidence from Italy. Available at SSRN 3634793.

Garrett, T.A. (2009) War and pestilence as labor market shocks: US manufacturing wage growth 1914–1919. *Economic Inquiry*, 47(4): 711–725.

Grusky, D.B., B. Western, and C. Wimer (2011) *The Great Recession*. New York: Russell Sage Foundation

Han, J., B.D. Meyer and J.X. Sullivan (2020) Income and Poverty in the COVID-19 Pandemic. NBER Working Paper No. 27729.

ILO (2020) COVID-19 and the world of work: Impact and policy responses. ILO Monitor 1st Edition. Geneva: ILO.

Immervoll, H., C. O'Donoghue and H. Sutherland (1999) An Introduction to EUROMOD. EUROMOD Working Paper EM0/99.

Immervoll H. and C. O'Donoghue (2004) What Difference Does a Job Make? The Income Consequences of Joblessness in Europe. In: D. Gallie (eds), *Resisting Marginalisation: Unemployment Experience and Social Policy in the European Union*, Oxford University Press, Oxford, 105-139.

Jenkins, S. P., A. Brandolini, J. Micklewright and B. Nolan (2013) *The Great Recession and the Distribution of Household Income*. Oxford University Press, Oxford.

Lustig, N., V.M. Pabon, F. Sanz and S.D. Younger (2020) The Impact of Covid-19 Lockdowns and Expanded Social Assistance on Inequality, Poverty and Mobility in Argentina, Brazil, Colombia and Mexico. CEQ Working Paper 92

O'Donoghue, C., Solognon, D.M., Kyzyma, I. and McHale, J. (2020) Modelling the Distributional Impact of the COVID-19 Crisis. IZA Discussion Paper No. 13235.

OECD (2020) Supporting people and companies to deal with the Covid-19 virus: options for an immediate employment and social-policy response. ELS Policy Brief on the Policy Response to the Covid-19 Crisis, OECD, Paris.

Patel, J.A., Nielsen, F.B.H., Badiani, A.A., Assi, S., Unidkat, V.A., Patel, B., Ravindrane, R. and Wardle, H. (2020) Poverty, inequality and COVID-19: the forgotten vulnerable. *Public Health*, 183: 110–111.

Petreski, M., Petreski, B., Tomovska Misoska, A., Gerovska Mitev, M., Parnardzieva Zmejkova, M., Dimkovski, V., Morgan, N. (2020) The Social and Economic Effects of Covid-19 on Children in North Macedonia: Rapid Analysis and Policy Proposals. Finance Think Policy Studies 2020-07/30, Finance Think - Economic Research and Policy Institute.

Petreski, M. and Jovanovic, B. (2016) Do Remittances Reduce Poverty and Inequality in the Western Balkans? Evidence from Macedonia? In: Giordano, C., Hayoz, N. and Herlth, J.(Eds.). (eds) *Diversity of Migration in South East Europe*. Fribourg, RRPP Joint Volume, p.85-109.

Petreski, B. and Petreski, M. (forthcoming 2020) Dynamic microsimulation modelling of potential pension reforms in North Macedonia. *Journal of Pension Economics and Finance*.

Salgado, M.F., F. Figari, H. Sutherland, A. Tumino (2014) Welfare Compensation for Unemployment in the Great Recession. *The Review of Income and Wealth,* 60(S1):S177-S204.

Sumner, A., C. Hoy and E. Ortiz-Juarez (2020) Estimates of the impact of COVID-19 on global poverty. WIDER Working Paper 2020/43.

Vos, R., W. Martin, and D. Laborde (2020) How much will global poverty increase because of COVID-19?. Downloaded at: https://www.ifpri.org/blog/how-much-will-global-povertyincrease-because-covid-19 [11 October 2020]

World Bank (2020) Regular Economic Report for the Western Balkans No. 17: Economic and Social Impact of Covid-19. Poverty and welfare of households.

# Annex – A general description of MK-MOD

MK-MOD is a tax and benefit microsimulation model for North Macedonia. It estimates the effects of changes in social and fiscal policies on measures of personal income and household welfare. MK-MOD is a national model, which implies that it provides perspective on policies implemented at national level, however strictly retaining the features of its patron, the EUROMOD. Hence, MK-MOD is suitable for integration into EUROMOD, so as to provide also comparisons where the same or similar policies have been implemented.

MK-MOD calculates household disposable income for each household in a representative set of micro-data, the Quality of Life Survey. Such calculation combines elements of gross income taken from the survey (or, more succinctly, the net income reported but grossed based on the current tax & benefit rules). Then, the current tax & benefit system is applied (the baseline), as well any policy change specified by the user.

To construct the pre-pandemic tax & benefit system of North Macedonia ("before" situation), MK-MOD relies on the following legal acts: Personal Income Tax Law, Social Contributions Law, Law on Employment and Insurance Against Unemployment, Social Protection Law, Child Protection Law, Law on Social Security of elderly, Family Law, Law on Pension and Disability Insurance. To construct the post-pandemic tax and benefit system ("after" situation), on top of the above laws, the following decrees were used: Decree with the power of law for application of the Social Protection Law during the state of emergency, Decree with the power of law for financial support of the employers from the private sector affected by the health-economic crisis caused by the virus COVID-19, for payment of salaries for the months of April and May 2020, Decree with the power of law for subsidizing the payment of compulsory social insurance contributions during a state of emergency, and Decree with the power of law for financial support of low-income citizens and employees, young people and health professionals by issuing a domestic payment card intended for purchase of Macedonian products and services during an emergency.

The first-round effect of any policy change is the simple arithmetic difference between the "before" and "after" situations and their associated calculations.

The policy areas for which changes can be simulated in this manner include income taxes, social contributions, pension benefits, social assistance benefits of all types (guaranteed minimum income, child allowances, educational allowances, one-time financial support, except in cases where it is tied to a very specific condition (e.g. deafness) which is not captured by the survey), some forms of property taxes and indirect taxes (despite, indirect taxed in the current form of MK-MOD are not simulated, but could be with some additional effort). These are the components of the tax-benefit system which are most commonly covered in national models. It is to note that MK-MOD allows for the work of the automatic stabilizers to the extent to which they are static in nature. For example, a decline in household market income automatically makes the household eligible for social assistance benefit (assuming that property conditions are also met). On the other hand, automatic stabilizers involving a dynamic component are not included in a static model, e.g. the unemployment insurance which is dependent on observation of the temporal transition from employment into unemployment.

The primary output from MK-MOD is the household disposable income. It comprises the following broad components: wage and salary income plus self-employment income plus property income plus other cash market income plus pension income plus cash benefit payments minus direct taxes and social insurance contributions. Hence, the output is a micro-level change in household disposable income as a consequence of the policy changes. This provides bases for the calculation of the aggregate effects on government revenues, distribution of gains and losses, differential effects on groups based on observable characteristics (gender, age, labor-market status, household type), effective marginal tax rates, replacement rates, the first-round impact on poverty and inequality and so on.

MK-MOD has its own limitations. Some caveats pertinent to this study have been elaborated in Section 3.1. Other caveats are of a general and predominantly technical nature. For example, the simulation power of MK-MOD is confined to the set of variables available in the underlying survey. For example, we are not able to simulate unemployment benefit, since majority of information determining its size (previous wage, tenure etc.) is not available in the survey. Likewise, disability pension is hard to simulate, because it depends on information about the occurrence of disability and previous incomes. For this and some other less important benefits, it is possible to simulate the amount, but not eligibility, and hence the latter is determined only on data showing benefit receipt (which is partial solution, as is confined on the existing recipients only).